\documentstyle[preprint,aps,epsf,floats]{revtex} 

\newcommand{\nn}{\nonumber} 
\newcommand{\ply}{ {\rm Li}_2 }

\newcommand{\lc}{\lowercase}

\begin{document}
\setlength\baselineskip{20pt}

\preprint{\tighten \vbox{\hbox{CALT-68-2227}
		\hbox{UTPT-99-10}
%		\hbox{nucl-th/9907xxx} 
}}

\title{The NN scattering $^3S_1-{}^3D_1$ mixing angle at NNLO}

\author{Sean Fleming\footnote{fleming@furbaide.physics.utoronto.ca}}
\address{\tighten Physics Department, University of Toronto, Toronto, Ontario,
M5S 1A7}
\author{Thomas Mehen\footnote{mehen@theory.caltech.edu} and 
Iain W.\ Stewart\footnote{iain@theory.caltech.edu} \\[4pt]}
\address{\tighten California Institute of Technology, Pasadena, CA 91125 }

\maketitle

{\tighten
\begin{abstract} 

The $^3S_1-{}^3D_1$ mixing angle for nucleon-nucleon scattering, $\epsilon_1$, is
calculated to next-to-next-to-leading order in an effective field theory with
perturbative pions.  Without pions, the low energy theory fits the observed
$\epsilon_1$ well for momenta less than $\sim 50$ MeV.  Including pions
perturbatively significantly improves the agreement with data for momenta up to
$\sim 150$ MeV with one less parameter.  Furthermore, for these momenta the
accuracy of our calculation is similar to an effective field theory calculation in
which the pion is treated non-perturbatively.  This gives phenomenological support
for a perturbative treatment of pions in low energy two-nucleon processes.  We
explain why it is necessary to perform spin and isospin traces in $d$ dimensions
when regulating divergences with dimensional regularization in higher partial wave
amplitudes.

\end{abstract}
}%end tighten
\vspace{0.7in}
%%]\narrowtext

\newpage

Effective field theory provides a technique for describing two-nucleon systems in
the most general way consistent with the symmetries of QCD\cite{orefs,ork}. In
Refs.~\cite{ksw1,ksw2}, Kaplan, Savage, and Wise (KSW) devised a power counting
that accounts for the effect of large scattering lengths.  With this power counting
the dimension six four-nucleon operators are non-perturbative, while pion exchange
and higher dimension operators are perturbative.  Powers of $a\,p$ are summed to
all orders ($p$ is a typical nucleon momentum, and $a$ is an S-wave scattering
length).  When pions are included in a manner consistent with chiral symmetry the
expansion is in powers of $Q/\Lambda$ where $Q=p$ or $m_\pi$, and $\Lambda$ is the
range of the theory.  For $p<m_\pi/2$ (below the pion cut), pions can be integrated
out leaving only contact interactions.  Therefore, the theory without pions is an
expansion in powers of $p/m_\pi$.  Note that for low enough momentum the theory
without pions will be more accurate since it is not limited by the additional
$m_\pi/\Lambda$ expansion.

A number of observables have been computed at next-to-leading order (NLO) with the
KSW power counting.  These include nucleon-nucleon phase shifts
\cite{ksw1,ksw2,epel}, Coulomb corrections to proton-proton scattering
\cite{kong2}, proton-proton fusion \cite{kong1}, electromagnetic form factors for
the deuteron \cite{ksw3}, deuteron polarizabilities \cite{chen}, $np\to d\gamma$
\cite{SSW}, Compton deuteron scattering \cite{chen2}, parity violating deuteron
processes \cite{Kaplan1}, and $\nu d \rightarrow \nu d$ \cite{butler}. Typically
errors are 30\%-40\% at leading order (LO) and of order 10\% at NLO indicating
$Q/\Lambda\sim 1/3$, or $\Lambda\sim 400\,{\rm MeV}$.  Since the expansion
parameter is fairly large, calculations at next-to-next-to-leading order (NNLO) are
necessary to achieve accuracy comparable to more conventional approaches.

In the KSW power counting the leading order diagrams for NN scattering are order
$1/Q$, so NNLO corresponds to an order $Q$ calculation.  In the theory without
pions, several of the observables listed above have been computed to NNLO
\cite{Mnopi}.   In the theory with pions the potential pion and local operator
contributions to the phase shift in the $^1S_0$ channel were calculated at NNLO in
Refs.~\cite{rupak,msconf}.  The deuteron quadrupole moment \cite{binger} has also
been computed at this order.  In this paper the $^3S_1-{}^3D_1$ mixing angle,
$\epsilon_1$, is calculated at NNLO in the theory with pions.  This calculation
provides a clear example of an observable for which the theory with perturbative
pions does better than the theory with only nucleons for momenta of order $m_\pi$,
and without additional parameters. In addition, for $p\sim m_\pi$ the accuracy of
this prediction is comparable to a calculation which treats the pion
nonperturbatively\cite{ork}.  

The relevant Lagrangian has terms with $0$, $1$, and $2$ nucleons:
\begin{eqnarray}  \label{Lpi}
{\cal L} &=& \frac{f^2}{8} {\rm Tr}\,( \partial^\mu\Sigma\: \partial_\mu 
\Sigma^\dagger )+\frac{f^2\omega}{4}\, {\rm Tr} (m_q \Sigma+m_q \Sigma^\dagger) 
+ N^\dagger \bigg( i D_0+\frac{\vec D^2}{2M} \bigg) N  \nn \\ 
&+& \frac{ig_A}2\, N^\dagger \sigma_i (\xi\partial_i\xi^\dagger -
\xi^\dagger\partial_i\xi) N -{C_0^{(^3S_1)}}
  {\cal O}_0^{(^3S_1)} +\frac{C_2^{(^3S_1)}}{8}{\cal O}_2^{(^3S_1)} 
  -{D_2^{(^3S_1)}} \omega {\rm Tr}(m^\xi ) {\cal O}_0^{(^3S_1)}  \nn\\[5pt]
 && -{C_2^{(SD)}} {\cal O}_2^{(SD)} + \ldots \,.
\end{eqnarray} 
Here $g_A=1.25$ is the nucleon axial-vector coupling, $\Sigma = \xi^2$, $f=131\,
{\rm MeV}$ is the pion decay constant, the chiral covariant derivative is
$D_\mu=\partial_\mu+\frac12 (\xi\partial_\mu\xi^\dagger +
\xi^\dagger\partial_\mu\xi)$, and $m^\xi=\frac12(\xi m_q \xi + \xi^\dagger m_q
\xi^\dagger)$, where $m_q={\rm diag}(m_u,m_d)$ is the quark mass matrix.  At the
order we are working $\omega {\rm Tr}(m^\xi )=w(m_u+m_d) = m_\pi^2=(137\,{\rm
MeV})^2$.  Eq.~(\ref{Lpi}) contains two-body nucleon operators 
\begin{eqnarray} \label{ops}
 {\cal O}_0^{(^3S_1)} &=& ( N^T P^{(^3S_1)}_i N)^\dagger ( N^T P^{(^3S_1)}_i N) 
     \nn \,, \\ 
 {\cal O}_2^{(^3S_1)} &=& ( N^T P^{(^3S_1)}_i N)^\dagger ( N^T
     P^{(^3S_1)}_i \:\tensor{\nabla}^{\,2} N) + h.c. \nn\,, \\
 {\cal O}_2^{(SD)} &=& ( N^T P^{(^3S_1)}_i N)^\dagger ( N^T
     P^{(^3D_1)}_i N) + h.c.  \,,
\end{eqnarray} 
where the projection matrices are
\begin{eqnarray} \label{proj}
    P_i^{({}^3\!S_1)} = { (i\sigma_2 \sigma_i  ) \, (i\tau_2) \over 2\sqrt{2} }\,,  \qquad
    P_i^{({}^3\!D_1)} =  \frac{n}{4\sqrt{n-1}} \: \Big( \tensor{\nabla}_i 
       \tensor{\nabla}_j -  \frac{\delta_{ij}}{n}\:\tensor{\nabla}^{\,2}  \Big) \: 
        P_j^{({}^3\!S_1)} \,,
\end{eqnarray}
$d=n+1$ is the space-time dimension, and $\tensor{\nabla} = \overleftarrow{\nabla}-
\overrightarrow{\nabla}$. The derivatives in Eqs.~(\ref{ops}) and (\ref{proj})
should really be chirally covariant, however, only the ordinary derivative is
needed for the calculation in this paper. $C_0^{(^3S_1)}$, $C_2^{(^3S_1)}$,
$D_2^{(^3S_1)}$, and $C_2^{(SD)}$ in Eq.~(\ref{Lpi}) are normalized so that the
on-shell Feynman rules in the center of mass frame are 
\begin{eqnarray}  \label{C2sd}
\\[-15pt]
  \begin{picture}(20,10)(1,1)
      \put(1,3){\line(1,1){10}} \put(1,3){\line(1,-1){10}} 
      \put(1,3){\line(-1,1){10}} \put(1,3){\line(-1,-1){10}}
      \put(-7,17){\mbox{\footnotesize $C_0^{(^3S_1)}$}}
      \put(-24,0){\mbox{\scriptsize ${}^3S_1$}}  
      \put(15,0){\mbox{\scriptsize ${}^3S_1$}} 
  \end{picture} 
   \quad &=&  -i\,C_0^{(^3S_1)}\,, \phantom{m_\pi^2\,} \qquad\qquad\qquad
  \begin{picture}(20,10)(1,1)
      \put(1,3){\line(1,1){10}} \put(1,3){\line(1,-1){10}} 
      \put(1,3){\line(-1,1){10}} \put(1,3){\line(-1,-1){10}}
      \put(-7,17){\mbox{\footnotesize $C_2^{(^3S_1)}$}}
      \put(-24,0){\mbox{\scriptsize ${}^3S_1$}}  
      \put(15,0){\mbox{\scriptsize ${}^3S_1$}} 
  \end{picture} 
   \quad =  -i\,C_2^{(^3S_1)}\, p^2 \,, \nn\\[20pt]
  \begin{picture}(20,10)(1,1)
      \put(1,3){\line(1,1){10}} \put(1,3){\line(1,-1){10}} 
      \put(1,3){\line(-1,1){10}} \put(1,3){\line(-1,-1){10}}
      \put(-7,17){\mbox{\footnotesize $D_2^{(^3S_1)}$}}
      \put(-24,0){\mbox{\scriptsize ${}^3S_1$}}  
      \put(15,0){\mbox{\scriptsize ${}^3S_1$}} 
  \end{picture} 
   \quad &=&  -i\,D_2^{(^3S_1)}\, m_\pi^2 \,, \qquad\qquad\qquad
  \begin{picture}(20,10)(1,1)
      \put(1,3){\line(1,1){10}} \put(1,3){\line(1,-1){10}} 
      \put(1,3){\line(-1,1){10}} \put(1,3){\line(-1,-1){10}}
      \put(-5,15){\mbox{\footnotesize $C_2^{(SD)}$}}
      \put(-24,0){\mbox{\scriptsize ${}^3S_1$}}  
      \put(15,0){\mbox{\scriptsize ${}^3D_1$}} 
  \end{picture} 
   \quad =  i\,C_2^{(SD)}\, p^2 \,, \nn
\end{eqnarray} 
where $p$ is the momentum of the nucleon.  From now on the superscript $(^3S_1)$
will be dropped.  Eq.~(\ref{C2sd}) is correct even if spin and isospin traces are
performed in $n$ dimensions.  

To regulate ultraviolet divergences it is convenient to use dimensional
regularization, which respects all the symmetries of the Lagrangian.  When using
dimensional regularization it is necessary to perform spin traces in $n$ dimensions
in order not to break rotational symmetry.  This is important for calculating
divergent graphs in higher partial waves.   For the nucleon theory it is convenient
to also continue the isospin traces to $n$ dimensions so that the regulator does
not break the Wigner symmetry \cite{Wigner} of the lowest order Lagrangian
\cite{msw}.  Spin and isospin polarization vectors are then normalized so that
\begin{eqnarray}
    \sum_i \,  \epsilon_i \epsilon_i^* = d-1 = n  \,.
\end{eqnarray}
For the scattering $NN(\epsilon_i)\to NN(\epsilon_j)$, $i=j$ so calculations may be 
simplified by setting
\begin{eqnarray}
     \epsilon_i \epsilon_j^* \to {\delta^{ij} \over  n}  \,.
\end{eqnarray}
A more detailed discussion of traces in $n$ dimensions is given in Appendix A.

To implement the KSW power counting it is useful to use a renormalization scheme
where the power counting is manifest, such as PDS\cite{ksw1,ksw2} or OS
\cite{ms0,ms1}. (In this paper the PDS scheme will be used.) In these schemes
coefficients of certain four-nucleon operators have power law dependence on the
renormalization point, $\mu_R$, and taking $\mu_R\sim p\sim m_\pi \sim Q$ makes
the power counting manifest.  The size of these coefficients is larger than naive
dimensional analysis would predict due to the presence of a non-trivial fixed point
for $a\to \infty$.  A consequence of this is that bubble graphs with $C_0$'s must be
summed to all orders. This sums all powers of $a\,p$\cite{ksw1,bira}.  The
$^3S_1$ coefficients in Eq.~(\ref{Lpi}) scale as $C_0(\mu_R)\sim 1/Q$,
$C_2(\mu_R)p^2\sim Q^0$, and $m_\pi^2 D_2(\mu_R)\sim Q^0$.  These parameters
are fixed by the $^3S_1$ phase shift at NLO.  $C_2^{(SD)}$ is an unknown parameter
and enters into the $^3S_1-{}^3D_1$ amplitude at order $Q$.  This is clear from the
beta function for $C_2^{(SD)}(\mu_R)$ in the theory without pions:
\begin{eqnarray}
   \beta_2^{(SD)} =\: \mu_R {\partial\over \partial\mu_R}\: C_2^{(SD)}(\mu_R)\: =\:
   \bigg( {M\mu_R\over 4\pi} \bigg) C_0^{({}^3\!S_1)}(\mu_R)
       \:C_2^{(SD)}(\mu_R)\,.
\end{eqnarray}
Solving this equation gives $p^2\,C_2^{(SD)}(\mu_R)\sim p^2/\mu_R\sim Q$.  As 
discussed below, pions give $C_2^{(SD)}$ an additional logarithmic dependence on 
$\mu_R$.  

The leading order $^3S_1-{}^3S_1$ amplitude is 
\begin{eqnarray}
   {\cal A}^{(-1)} = -\frac{4\pi}{M} \frac1{\gamma+ip}\,,\qquad\quad 
 	\gamma=\frac{4\pi}{MC_0}+\mu_R  \,.
\end{eqnarray}
This amplitude has a pole at $p=i\gamma$ corresponding to the deuteron bound
state. The deuteron has binding energy $B=2.22\,{\rm MeV}$, so $\gamma=
\sqrt{M B} = 45.7\,{\rm MeV}$.  With this boundary condition the difference between 
$\gamma$ and the observed scattering length $a$ is obtained from perturbative 
contributions to $C_0$\cite{ms0}
\begin{eqnarray} \label{C0expn}
  C_0(\mu_R) = C_0^{np}(\mu_R) + C_0^{(0)}(\mu_R) + \ldots \,,
\end{eqnarray}
where $C_0^{(0)}(\mu_R)\sim Q^0$.  In the PDS scheme the expansion in
Eq.~(\ref{C0expn}) is necessary to obtain $\mu_R$ independent amplitudes at each
order in $Q$.  This expansion is also necessary to ensure that higher order 
corrections do not give an amplitude with spurious higher order poles\cite{ms0,ms1}.

The S matrix for the $^3S_1$ and $^3D_1$ channels is $2 \times 2$ and
can be parameterized using the convention in Ref.~\cite{Stapp} :
\begin{eqnarray}  \label{S}
  S = {\mathbf{1}} + \frac{i\, M p}{2\pi} \left( \begin{array}{cc} {\cal A}^{SS} & 
      {\cal A}^{SD} \\ {\cal A}^{SD} & {\cal A}^{DD} \end{array} \right) 
   =   \left(  \begin{array}{cc}
       e^{2i\bar\delta_0} \cos 2\bar\epsilon_1 & i\, e^{i\bar\delta_0+i\bar\delta_2} 
       \sin 2\bar\epsilon_1 \\
       i\, e^{i\bar\delta_0+i\bar\delta_2} \sin 2\bar\epsilon_1 & e^{2i\bar\delta_2} 
       \cos 2\bar\epsilon_1 
     \end{array}   \right) \,.
\end{eqnarray}
In this parameterization the mixing angle is given by
\begin{eqnarray} \label{fullep1}
  \sin(2\,\bar \epsilon_1) = \frac{M p}{2\pi} \frac{ {\cal A}^{SD} } 
  { \sqrt{ \Big[1+\frac{ip M}{2\pi} {\cal A}^{SS}\Big] 
     \Big[1+\frac{ip M}{2\pi} {\cal A}^{DD}\Big] + \Big(\frac{Mp}{2\pi}\Big)^2
   [ {\cal A}^{SD} ]^2  } } \,.
\end{eqnarray}
The phase shifts and mixing angle can be expanded in powers of $Q/\Lambda$  
\begin{eqnarray}  \label{dexpn}
   \bar\delta_0 = \bar\delta_0^{(0)}  + \bar\delta_0^{(1)} + \ldots \,,\qquad 
   \bar\delta_2 = 0  + \bar\delta_2^{(1)} + \ldots \,,\qquad 
   \bar\epsilon_1 = 0 + \bar\epsilon_1^{(1)} + \bar\epsilon_1^{(2)} + \ldots \,,
\end{eqnarray}
where the superscript denotes the order in the $Q$ expansion.  The phase shifts and
mixing angles start at one higher order in $Q$ than the amplitudes because of the
factor of $p$ in Eq.~(\ref{S}).  Since ${\cal A}^{SD}$ starts at $Q^0$, there is no order
$Q^0$ contribution to $\bar\epsilon_1$.  This is consistent with the fact that this
angle is much smaller than the ${}^3\!S_1$ phase shift.  In the PDS scheme,
expressions for $\bar\delta_0^{(0,1)}$, $\bar\delta_2^{(1)}$, and $\bar\epsilon_1^{(1)}$
were given in Ref.~\cite{ksw2}. 
\begin{figure}[!t]
  \centerline{\epsfxsize=16.truecm \epsfysize=5.truecm\epsfbox{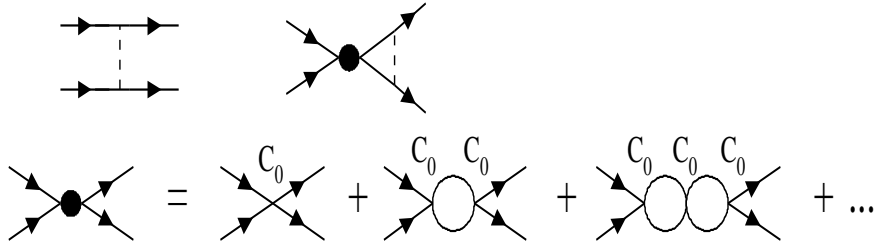}  }
{\tighten \caption[1]{The two order $Q^0$ diagrams that contribute to 
$\bar\epsilon_1$ \cite{ksw2}.   The solid lines are nucleons and the dashed lines are 
potential pions.} \label{diag_lo} }   
\end{figure}  
Our main result is the calculation of $\bar\epsilon_1^{(2)}$. The NNLO predictions
for $\bar\delta_0^{(2)}$ and $\bar\delta_2^{(2)}$ are not needed to calculate
$\bar\epsilon_1^{(2)}$ and will be presented in a future publication \cite{fms2}. 
Expanding both sides of Eq.~(\ref{fullep1}) in powers of $Q$
gives\footnote{\tighten The branch cut for the square root in Eq.~(\ref{epss}) is
taken to be on the positive real axis.  This is consistent with $\bar \delta_0(p\to
0)=\pi$. The sign of our $^3D_1$ state is the opposite of Ref.~\cite{ksw2}, making
$A^{SD(0)}$ in Eq.~(\ref{Q0}) have the opposite overall sign.}
\begin{eqnarray}  \label{epss}
   \bar\epsilon_1^{(1)} &=&  \frac{M p}{4\pi}  \frac{ {\cal A}^{SD(0)} }
        {\Big[1+2\frac{ip M}{4\pi} {\cal A}^{(-1)}\Big]^{1/2} }     
     = \frac{M p}{4\pi} \ \Big| {\cal A}^{(-1)}\Big| \ 
        \frac{ {\cal A}^{{SD}(0)}}{ {\cal A}^{(-1)}}    \,, \\
   \bar\epsilon_1^{(2)} &=& \frac{M p}{4\pi}  \frac{ {\cal A}^{{SD}(1)} }
        {\Big[1+2\frac{ip M}{4\pi} {\cal A}^{(-1)}\Big]^{1/2} } 
        -i \epsilon_1^{(1)}\Big[ \delta_0^{(1)} +  \delta_2^{(1)} \Big]  
     = \frac{M p}{4\pi} \ \Big| {\cal A}^{(-1)}\Big| \ {\rm Re}\Big[ 
        \frac{ {\cal A}^{{SD}(1)}}{ {\cal A}^{(-1)}}  \Big]  \,. \nn
\end{eqnarray}
$\bar\epsilon_1^{(1)}$ is determined by the order $Q^0$ graphs in Fig.~\ref{diag_lo}
and does not involve any free parameters. The order $Q^0$  mixing amplitude is 
\cite{ksw2}
\begin{eqnarray}  \label{Q0}
  {\cal A}^{SD(0)} &=& \sqrt{2} \frac{M g_A^2}{8\pi f^2}\: {\cal A}^{(-1)} 
       \bigg\{ m_\pi \, {\rm Re}[{\cal X(\alpha)}] - \frac{\gamma}{\alpha} 
       {\rm Im}[{\cal X(\alpha)}] \bigg\} \,, \\
   {\cal X}(\alpha) &=& -\frac{3}{4\alpha^2}-\frac{3i}{4\alpha}+\frac{i\alpha}2 + 
     i \bigg( \frac1{2\alpha} + \frac3{8\alpha^3} \bigg) \ln(1-2 i\alpha) \,, \nn
\end{eqnarray}
where
\begin{eqnarray}
  \alpha &\equiv & \frac{p}{m_\pi} \,.
\end{eqnarray}

\begin{figure}[!t]
  \centerline{\epsfxsize=18.truecm \epsfysize=7.truecm\epsfbox{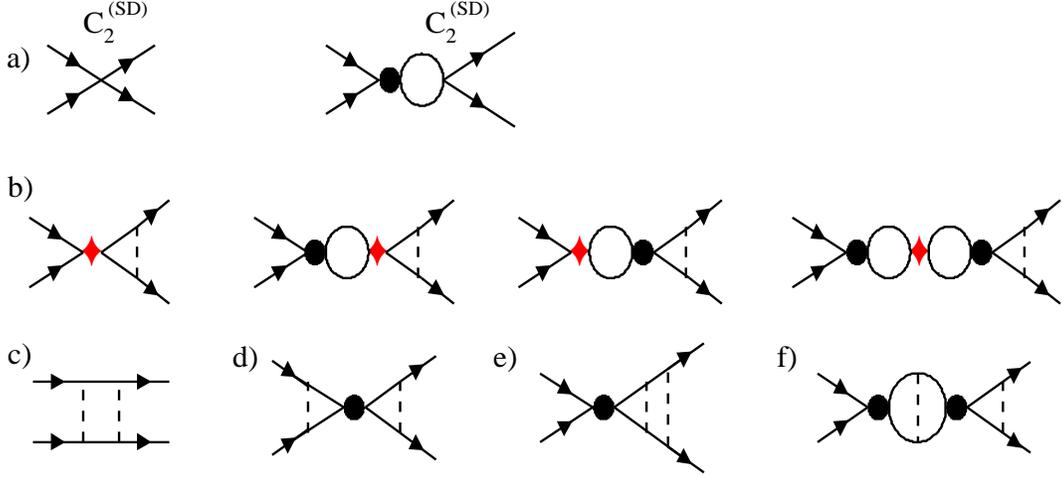}  }
{\tighten \caption[1]{Order $Q$ diagrams for $\bar\epsilon_1$.  The filled circle is 
defined in Fig.~\ref{diag_lo}, and the diamonds in b) denote insertions of the 
$^3S_1-{}^3S_1$ operators with coefficients $C_0^{(0)}$, $C_2$ or $D_2$. }  
\label{diag_Q}  } 
\end{figure}

At order $Q$, the Feynman diagrams that contribute to the $^3S_1-{}^3D_1$ amplitude
are shown in Fig.~\ref{diag_Q}.  In addition to potential pions, at this order the
S-wave phase shifts can have contributions from diagrams with radiation pions
\cite{ksw2}.  Performing the energy loop integrals using contour integration,
potential pions occur when a pole from a nucleon propagator is taken.  Radiation
pion contributions come from taking a pole in a pion propagator.  For graphs with
radiation pions it is necessary to count powers of $p\sim Q_r=\sqrt{M\,m_\pi}$
\cite{ms2} and then scale down to $p\sim m_\pi$. Order $Q$ contributions can come
from $Q_r^3$ and $Q_r^4$ radiation pion graphs\cite{msconf}, however these vanish
for a $^3S_1-{}^3D_1$ transition.   Soft pion graphs begin at order $Q_r^2$, and
for $p\sim m_\pi$ are order $Q^2$ \cite{ms2}.  Relativistic corrections begin at
order $Q^2$ and therefore are not included.
 
In dimensional regularization a graph with $k$ loops includes a factor of
$(\mu_R/2)^{k(4-d)}$ (where the extra $2$ is inserted for convenience).  Spin and
isospin traces will be evaluated in $d-1$ dimensions for the reasons discussed in
Appendix A.  Of the graphs in Fig.~\ref{diag_Q} only e) and f) are divergent in
$d=4-2\epsilon$ dimensions.  The divergence in f) is cancelled by a graph with the
NLO $\delta^{\rm uv}D_2$ counterterm\footnote{ \tighten The bare coefficients in
Eq.~(\ref{Lpi}) are written as $C^{\rm bare}=\delta^{\rm uv}C + C^{\rm finite}$. In PDS
additional finite subtractions are made so that $C^{\rm finite}= C(\mu_R)-\sum
\delta^n C(\mu_R)$, see Ref.~\cite{ms1}.} given by Eq.~(5.2) of Ref.~\cite{ms1}.  The
$p^2/\epsilon$ divergence in e) is cancelled by the new counterterm 
\begin{eqnarray}
  \delta^{\rm uv}C_2^{(SD)} = {3\sqrt{2} C_0^{\rm finite} \over 10} \Big( 
   \frac{M g_A^2}{8\pi f^2}\Big)^2 \Big( \frac{1}{2\epsilon} -\gamma_E +
   \ln\pi \Big) \,.
\end{eqnarray}
Note that it is crucial to indicate what constants are subtracted along with the
$1/\epsilon$ pole.  The coupling $C_2^{(SD)}$ is determined from a fit to the
observed $\bar\epsilon_1$.  If the extracted value is to be used in other
calculations, then its exact definition including finite
subtractions will be needed\footnote{\tighten We have not compared our value
of $C_2^{(SD)}(m_\pi)$ to the value extracted from the deuteron quadrupole
moment \cite{binger} for this reason.}.  The divergence in Fig.~\ref{diag_Q}\,e) 
induces $\ln(\mu_R)$ dependence in $C_2^{(SD)}(\mu_R)$.  In PDS
\begin{eqnarray} \label{C2pi}
  C_2^{(SD)}(\mu_R) = \kappa C_0(\mu_R) - { 3\sqrt{2} \over 10} C_0(\mu_R)
    \bigg({M g_A^2 \over 8\pi f^2 } \bigg)^2 \ln\Big({\mu_R^2 \over \lambda^2}
    \Big)\,,
\end{eqnarray}
where $\kappa$ and $\lambda$ are constants.  Note that there is only one 
unknown in Eq.~(\ref{C2pi}) since a shift in the value of $\kappa$ can be 
compensated by changing the value of $\lambda$.

\begin{figure}[!t]
  \centerline{\epsfxsize=17.truecm \epsfysize=10.truecm\epsfbox{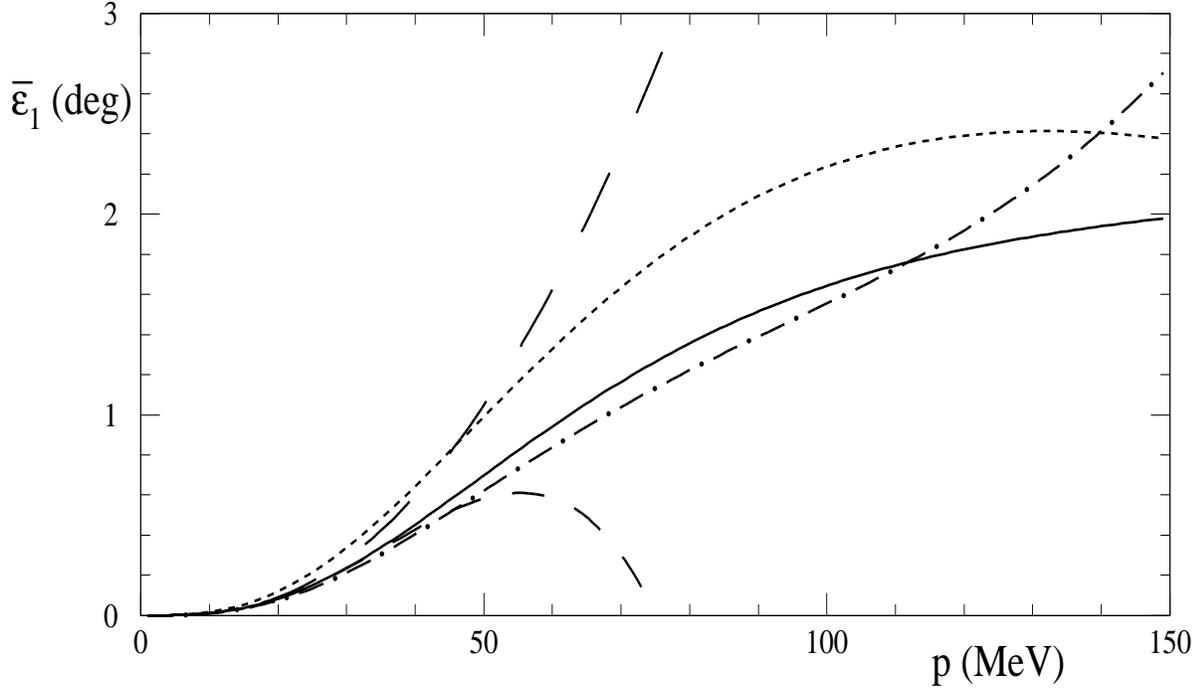}  }
{\tighten \caption[1]{Predictions for the $^3S_1-{}^3D_1$ mixing parameter
$\bar\epsilon_1$.  The solid line is the multi-energy Nijmegen partial wave analysis
\cite{NijPW}.  The long and short dashed lines are the order $Q^2$ and $Q^4$ 
predictions in the theory without pions \cite{Mnopi}.  The dotted line is the order 
$Q$ prediction in the theory with pions from Ref.~\cite{ksw2}.  The dash-dotted 
line is the order $Q^2$ prediction in the theory with pions.}  \label{ep1a} }
\end{figure}

At order $Q$ the diagrams in Fig.~\ref{diag_Q} give the following amplitudes in 
the PDS scheme
\begin{eqnarray} 
  {\cal A}^{{SD}(1)} =  {\cal A}_a +  {\cal A}_b +  {\cal A}_c +  {\cal A}_d 
     +  {\cal A}_e + {\cal A}_f \,,
\end{eqnarray}
where 
\begin{eqnarray}  \label{AmpQ}
 {i\cal A}_a  &=&  i \, C_2^{(SD)}\, p^2 \bigg[ 1+ \frac{M {\cal A}^{(-1)}}{4\pi} 
    (ip+\mu_R) \bigg]  = -i \, {\cal A}^{(-1)}\, \frac{C_2^{(SD)}\, p^2}{C_0}   \,, \\
 {i\cal A}_b  &=& -i\, [ {\cal A}^{(-1)}]^2\:  \sqrt{2}\ \frac{(C_2\, p^2 +D_2\, m_\pi^2 + 
   C_0^{(0)} )}{C_0^2} \: \frac{m_\pi M g_A^2}{8\pi f^2} \: {\cal X}(\alpha) \,,   \nn\\
 {i\cal A}_c  &=&  i {3\sqrt{2}\over 2}\ \frac{M}{4\pi} \Big(\frac{g_A^2}{2 f^2} \Big)^2
   m_\pi \: {\cal Y}(\alpha) \,, \nn\\ 
{i\cal A}_d  &=& -i\,{\cal A}^{(-1)} \sqrt{2}\ \Big( \frac{M g_A^2}{8\pi f^2} \Big)^2 \,
      m_\pi^2\bigg[ i\alpha- \frac{i}{2\alpha}\ln(1-2i\alpha)\bigg]{\cal X}(\alpha)\,,\nn\\ 
 {i\cal A}_e  &=& i \,{\cal A}^{(-1)} \sqrt{2}\ \Big( \frac{M g_A^2}{8\pi f^2} \Big)^2 
      m_\pi^2 \bigg[  -\frac{3\alpha^2}{10} \ln\Big(\frac{\mu_R^2}{m_\pi^2}\Big) 
      -i\,\alpha\, {\cal X}(\alpha)   + {\cal Z}(\alpha)    \bigg] \nn \,, \\
 {i\cal A}_f  &=& -i [{\cal A}^{(-1)}]^2 \sqrt{2}\ \Big(\frac{M}{4\pi}\Big)^3 
      \Big(\frac{g_A^2}{2 f^2}\Big)^2 m_\pi^3 \bigg[(i\alpha)^2-\frac{\mu_R^2}{m_\pi^2}
      -\frac12 \ln\Big(\frac{\mu_R^2}{m_\pi^2}\Big) + \ln(1-2i\alpha) \bigg] 
      {\cal X}(\alpha) \,. \nn
\end{eqnarray}
The function ${\cal X}(\alpha)$ is given in Eq.~(\ref{Q0}), and the functions 
${\cal Y}(\alpha)$ and ${\cal Z}(\alpha)$ are given in Appendix B.
The sum of the amplitudes in Eq.~(\ref{AmpQ}) is: 
\begin{eqnarray}  \label{Asum}
 {\cal A}^{{SD}(1)} &=&  -{\cal A}^{(-1)}\ \zeta_6\: \alpha^2 - [{\cal A}^{(-1)}]^2 
   \sqrt{2}\: \frac{m_\pi^3 M g_A^2}{8\pi f^2} \: {\cal X}(\alpha)\ ( \zeta_1 \alpha^2 + 
   \zeta_2 )\\
   && + \sqrt{2} \frac{M m_\pi}{4\pi} \Big( \frac{g_A^2}{2 f^2} \Big)^2 \Bigg\{ 
     \frac{M m_\pi  {\cal A}^{(-1)}}{4\pi} \bigg[ 
    {\cal Z}(\alpha) + \frac{i}{2\alpha} \ln(1-2 i \alpha) {\cal X}(\alpha) \bigg] \nn\\
   && \qquad\qquad\qquad\qquad\quad  - \bigg[ \frac{M m_\pi  {\cal A}^{(-1)}}{4\pi}
          \bigg]^2 \ln(1-2i\alpha) {\cal X}(\alpha) + {3 \over 2}{\cal Y}(\alpha) + 
         {\cal X}(\alpha)  \Bigg\} \,, \nn
\end{eqnarray}
where $\zeta_1$, $\zeta_2$, and $\zeta_6$ are $\mu_R$ independent dimensionless 
combinations of coupling constants:
\begin{eqnarray} \label{zetas}
  \zeta_1 &=& {C_2(\mu_R) \over C_0(\mu_R)^2 } \,, \qquad
  \zeta_2 = {D_2(\mu_R) \over C_0(\mu_R)^2 } + {C_0^{(0)}(\mu_R) \over m_\pi^2
   C_0(\mu_R)^2} - {g_A^2 \over 2f^2} \bigg( \frac{M}{4\pi} \bigg)^2 \bigg[ \frac12 
    \ln\Big(\frac{\mu_R^2}{m_\pi^2}\Big) + {\mu_R^2 -\gamma^2 \over m_\pi^2} 
    \bigg]\,, \nn\\[10pt] 
 \zeta_6 &=& { m_\pi^2 C_2^{(SD)}(\mu_R) \over C_0(\mu_R)} + { 3\sqrt{2} \over 10}
    \bigg({M m_\pi g_A^2 \over 8\pi f^2 } \bigg)^2 \ln\Big({\mu_R^2 \over m_\pi^2}
    \Big)\,.
\end{eqnarray}
$\zeta_1$ and $\zeta_2$ also appear in the NLO $^3S_1$ amplitude (see
Eq.~(\ref{Samp})). $\zeta_2$ can be eliminated by imposing the condition that no
spurious double pole should appear in this amplitude\cite{msconf}:
\begin{eqnarray}\label{gfc}
   \zeta_2 = {\gamma^2 \over m_\pi^2} \zeta_1 -  {g_A^2 \over 2 f^2} 
     \left(M \over 4 \pi \right)^2 {\rm log}\left(1 ~+ {2 \gamma \over m_\pi} 
   \right).
\end{eqnarray}  
The constant $\zeta_1$ is extracted from a fit to the $^3S_1$ phase shift at NLO. 
The order $Q$ contribution to $\bar\epsilon_1$ contains one unknown parameter,
$\zeta_6$ or $C_2^{(SD)}(\mu_R)$.  This parameter is determined by fitting to the
value of $\bar\epsilon_1$ from the Nijmegen partial wave analysis\cite{NijPW} at
low momentum.  Results for $\bar\epsilon_1$ are shown in Fig.~\ref{ep1a}. The solid
line is the Nijmegen result. The order $Q$ result in the theory with pions
\cite{ksw2} is shown by the dotted line. The result of the order $Q^2$ calculation
in the theory with pions is given by the dot-dashed line in Fig.~\ref{ep1a}.  
The values used in Fig.~\ref{ep1a} are
\begin{eqnarray} \label{nums}
  \gamma=45.7\,{\rm MeV}\,, \qquad \zeta_1 = 0.2345\,, \qquad \zeta_2=  -0.1038\,, 
  \qquad \zeta_6= 0.385 \,.
\end{eqnarray}
The value of $\zeta_6$ in Eq.~(\ref{nums}) corresponds to 
\begin{eqnarray}
 C_2^{(SD)}(m_\pi) = -4.56\:{\rm fm^{4}} \,.
\end{eqnarray}

For comparison results have also been shown in Fig.~\ref{ep1a} for the theory
without pions \cite{Mnopi}, where the prediction for $\bar\epsilon_1$ begins at
order $Q^2$.  The long dashed line is the order $Q^2$ result and the theory
prediction has one free parameter.  The short dashed line is the order $Q^4$ result
which has two free parameters.  With one less free parameter, the order $Q^2$
prediction of the theory with pions does better than the order $Q^4$ prediction of
the theory without pions for $p>50\,{\rm MeV}$.  In fact the theory without pions
breaks down around $m_\pi/2$, as expected since this is where the pion cut begins. 
It has been noted in the literature \cite{Cohen3} that many observables may not
test the power counting for perturbative pions.  As can be seen from
Fig.~\ref{ep1a}, the mixing parameter provides an example in which perturbative
pions clearly give improved agreement with the data.  

The dot-dashed line in Fig.~\ref{ep1a} improves over the order $Q$ result for
$p<140\,{\rm MeV}$.  For $p\sim m_\pi$, the error in the order $Q^2$ prediction for
$\bar\epsilon_1$ is $\sim 20\%$.  Recall that the mixing angle is small and an
error of $\sim 0.5^\circ$ is consistent with our expectation for a NNLO
calculation.  It is interesting to ask how sensitive the results in Fig.~\ref{ep1a}
are to the choice of parameters.  If we use the $^3S_1$ scattering length to fix
$\gamma$ instead of the deuteron binding energy then the order $Q^0$ result (dotted
line) increases by $\sim 1^\circ$ for $p\sim m_\pi$.  Therefore, the mixing angle
is quite sensitive to the location of the pole.  On the other hand, the NNLO
prediction is not sensitive to the value of $\zeta_1$ obtained from fitting the
$^3S_1$ phase shift.  This is because $\bar\epsilon_1^{(2)}$ in Eq.~(\ref{epss})
depends on the linear combination
\begin{eqnarray}
  z = \zeta_6 - 0.56\, \zeta_1 \,,
\end{eqnarray}
but is insensitive to the orthogonal combination.  A change in $\zeta_1$ can be
compensated by a change in $\zeta_6$ while keeping $z \simeq 0.255$.  Solutions
with the same $z$ give similar predictions,  for instance, taking $\zeta_1=0.300$
and $\zeta_6=0.423$ gives an order $Q^2$ phase shift that differs by $< 0.08^\circ$
from the one shown in Fig.~\ref{ep1a}.  

A further test of the convergence of the $Q$ expansion is provided by examining the
extent to which the amplitude violates unitarity. When Eq.~(\ref{fullep1}) is
expanded in powers of $Q$ the expression for $\bar \epsilon_1$ is explicitly real
at each order in $Q$.  However, one could insert the NLO expression for ${\cal
A}^{SS}$ and ${\cal A}^{DD}$ and the NNLO expressions for ${\cal A}^{SD}$ into
Eq.~(\ref{fullep1}) and solve for $\bar \epsilon_1$ without making a $Q$ expansion.
The resulting $\bar \epsilon_1$ will have an imaginary part which is order $Q^3$ in
the power counting.  Comparing the imaginary part of $\bar \epsilon_1$ calculated
using Eq.~(\ref{fullep1})  to $\bar \epsilon_1^{(1)}+\bar \epsilon_1^{(2)}$ gives
$| {\rm Im}(\bar \epsilon_1)/ (\bar\epsilon_1^{(1)}+ \bar \epsilon_1^{(2)})| \le
0.2$ for $p \le 180\,{\rm MeV}$, which is of the expected size for an order $Q^2$
quantity.  Also, for $p\le m_\pi$ the ratio $|{\cal A}^{SD(1)}/{\cal A}^{SD(0)}|
\le 0.6$, which is consistent with an expansion parameter of order $1/2$.  The
agreement of the size of these terms with our expectations suggests that the $Q$
expansion is under control. 

\begin{figure}[!t] 
  \centerline{\epsfxsize=15.truecm \epsfysize=8.truecm\epsfbox{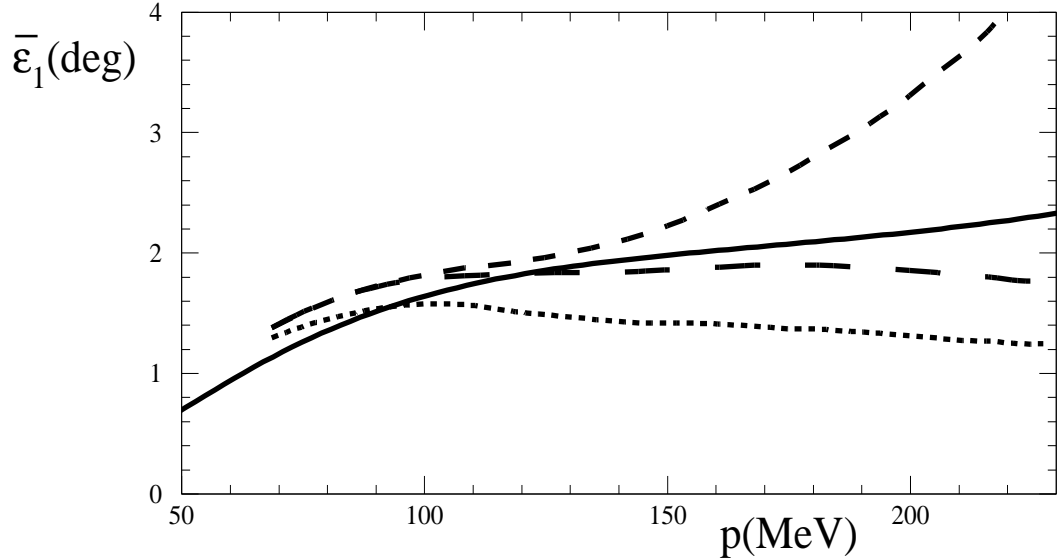}  } 
{\tighten \caption[1]{Prediction for $\bar\epsilon_1$ from Ref.~\cite{ork}. The fit
was done to the partial wave analysis in Ref.~\cite{NijPW} shown by the solid line.
The long dashed line uses the cutoff $\Lambda=0.6\, m_\rho$, the short dashed line 
uses $\Lambda= m_\rho$, and the dotted line uses $\Lambda=1.3\, m_\rho$.}
\label{diag_ork}} 
\end{figure}
In Ref.~\cite{ork}, the mixing angle is calculated using Weinberg's power counting.
In this approach, momentum power counting is applied to the potential and then the
Schroedinger equation is solved numerically.  Solving the Schroedinger equation
with the one pion exchange potential is equivalent to summing ladder graphs with
potential pion exchange to all orders. However, all necessary counterterms are not
included, so there is a residual dependence on the cutoff. This cutoff dependence
can be used to give an estimate of the uncertainty in the theoretical prediction
due to higher order effects.  We will compare our calculation with that of
Ref.~\cite{ork}, however it is important to keep in mind that Ref.~\cite{ork}
includes graphs which are higher order in $Q$ than those in Fig.~\ref{diag_Q}. 
Ref.~\cite{ork} also includes $\Delta$'s and more parameters are varied in the fit.
The results of Ref.\cite{ork} are shown in Fig.~\ref{diag_ork}.  Varying the cutoff
between $0.6\, m_\rho$ and $1.3\, m_\rho$ gives an uncertainty of $0.7^{\circ}$ at
$p = m_\pi$.  This uncertainty is comparable to the error in our fit which differs
from the data by $0.5^{\circ}$ at $p = m_\pi$.  The error in our calculation
increases for larger values of $p$ because our prediction grows with $p$ faster
than the observed $\bar \epsilon_1$. For these values of $p$ the nonperturbative
calculation suffers from considerable uncertainty.  For a cutoff equal to $m_\rho$,
the prediction grows with $p$, but with a lower value of the cutoff ($0.6\,
m_\rho$) the calculated $\bar\epsilon_1$ provides better agreement with data.  It
would be interesting to work to one higher order in $Q$ and/or include $\Delta$'s
with the KSW power counting to see if the agreement with data at higher $p$
improves.  At one higher order in $Q$ a four derivative four nucleon
$^3S_1-{}^3D_1$ operator appears.  However, using the renormalization group its
coefficient is determined in terms of $C_0$, $C_2$, and $C_2^{(SD)}$.

For momenta $p\ll m_\pi$, effective range expansions can be constructed for the
phase shifts and mixing angle. By integrating the pion out of the effective field
theory coefficients in this expansion can be predicted. In Ref.~\cite{Cohen2}
coefficients in the expansions of $p\,\cot\delta^{(^1S_0)}$, $p\,\cot\bar\delta_0$,
and $\bar\epsilon_1$ are obtained from the order $Q^0$ calculations in
Ref.~\cite{ksw2}.  Ref.~\cite{Cohen2} found that the effective field theory gives
parameter free predictions for the higher coefficients, but these did not agree
with fits \cite{Nij2} to the partial wave data.  However, it is not clear whether
the extraction of higher order terms in the expansion is accurate enough to test
the effective field theory \cite{ms0}.  In toy models it has been shown that the
convergence of the effective field theory predictions for these coefficients is
slow\cite{kap}.  This also seems to be the case when the effective field theory is
applied to real data.  In Ref.~\cite{msconf} it was found that the order $Q$
corrections to the coefficients of $p\,\cot\delta^{(^1S_0)}$ improve the agreement
with the fit values, however the observed convergence is rather slow.  

From the amplitude in Eq.~(\ref{Asum}) the order $Q^2$ corrections to the
momentum expansion of $\bar\epsilon_1$ can be derived. The expansion in the 
theory without pions takes the form \cite{Mnopi}
\begin{eqnarray}
   \bar\epsilon_1 = b_1 {p^3 \over \sqrt{p^2+\gamma^2} } + 
	b_2 {p^5 \over \sqrt{p^2+\gamma^2}} + \ldots  \,,
\end{eqnarray}
where $b_1$ and $b_2$ are constants.  $\bar\epsilon_1$
has a cut at $p=\pm i\gamma$, so the momentum expansion of
$\bar\epsilon_1$ only converges for $p<\gamma$. Clearly it would be more useful to
expand a function with better analyticity properties.  Following Ref.~\cite{Blatt} 
this can be done by parameterizing the S-matrix as:
\begin{eqnarray} \label{Blatte}
 S  = \left(  \begin{array}{cc} \cos \epsilon_1 & -\sin\epsilon_1 \\
            \sin\epsilon_1 & \cos \epsilon_1 \end{array} \right) 
        \left(  \begin{array}{cc} e^{2i\delta_0} & 0 \\ 0 & e^{2i\delta_2} 
   \end{array}\right)
        \left(  \begin{array}{cc} \cos \epsilon_1 & \sin\epsilon_1 \\
            -\sin\epsilon_1 & \cos \epsilon_1 \end{array}   \right) \,.
\end{eqnarray}
$p\cot\delta_0,\,$ $p^5\cot\delta_2,\,$ and $\epsilon_1$ have momentum expansions
with radius of convergence $m_\pi/2$ rather than $\gamma$.  For low energy
expansions these variables should be used.  The expressions for $\delta_{0,2}$ and 
$\bar\delta_{0,2}$ are the same to order $Q$.  The mixing angle in this 
parameterization is related to the one in Eq.~(\ref{S}) by
\begin{eqnarray}
  \tan(2\,\epsilon_1) = {\tan(2\, \bar\epsilon_1) \over \sin(\bar \delta_0- \bar 
  \delta_2)} = { 2 {\cal A}^{SD} \over {\cal A}^{SS}-{\cal A}^{DD} }\,.
\end{eqnarray}
In terms of the amplitudes, the first two terms in the $Q$ expansion of $\epsilon_1$ 
are
\begin{eqnarray}
  \epsilon_1^{(1)} &=& { {\cal A}^{SD(0)} \over {\cal A}^{(-1)} } \,, \qquad
  \epsilon_1^{(2)} =  {\rm Re}\bigg[ \frac{ {\cal A}^{{SD}(1)}}{ {\cal A}^{(-1)}}
  \bigg]- \frac{M\gamma}{4\pi}\: \epsilon_1^{(1)}\: \bigg[ {\cal A}^{DD(0)} - 
      |{\cal A}^{(-1)}|^2  { {\cal A}^{SS(0)} \over ({\cal A}^{(-1)})^2 } \bigg] 
   \ \,.
\end{eqnarray}

\begin{figure}[!t]
  \centerline{\epsfxsize=17.truecm \epsfysize=10.truecm\epsfbox{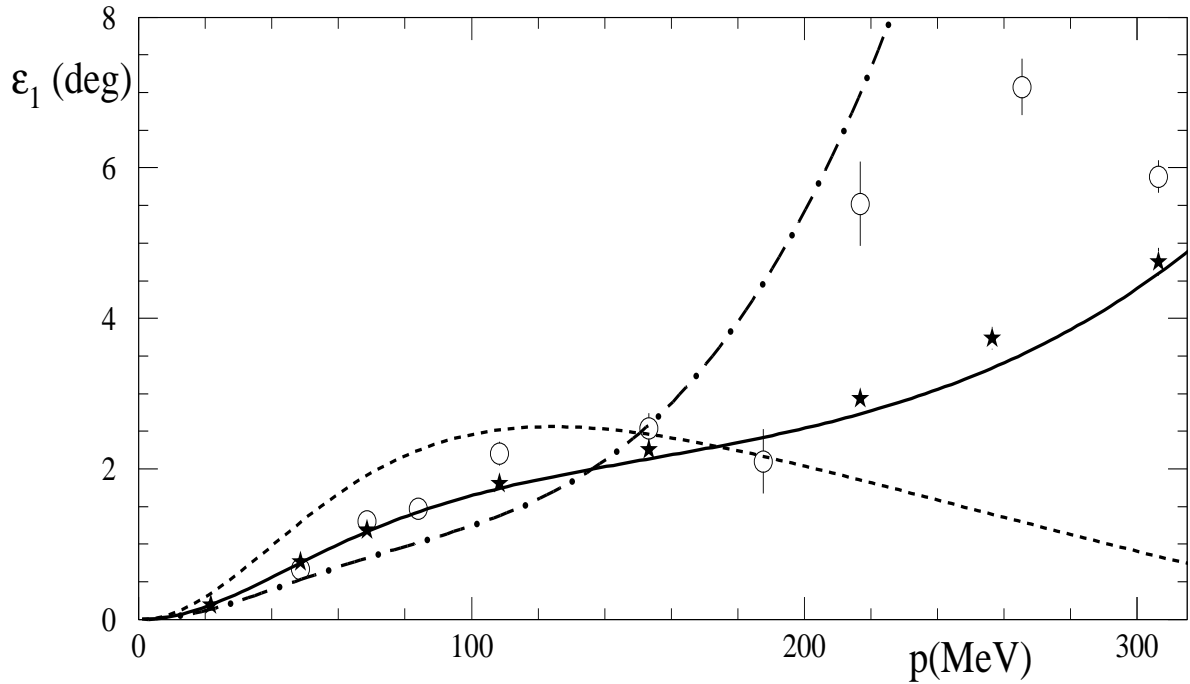}  }
{\tighten \caption[1]{Predictions for the mixing parameter
$\epsilon_1$ defined in Eq.~(\ref{Blatte}).  The solid line is the multi-energy 
Nijmegen partial wave analysis \cite{NijPW}.   The dotted line is the NLO prediction in 
the theory with pions from Ref.~\cite{ksw2}.  The dash-dotted line is the NNLO
prediction in the theory with pions.  The open circles are data from Virginia Tech
\cite{vpiPWA} and the stars are Nijmegen single energy data \cite{NijPW}
whose quoted errors are invisible on the scale shown.  }  \label{ep1b} }
\end{figure}
In Fig.~\ref{ep1b} we plot the order $Q$ and $Q^2$ effective field theory predictions
for $\epsilon_1$ using the parameters in Eq.~(\ref{nums}).  The open circles in
Fig.~\ref{ep1a} are data from Virginia Tech \cite{vpiPWA}.  The stars are the
Nijmegen single energy fit to the data \cite{NijPW} whose quoted errors are invisible
on the scale shown.  It seems somewhat strange that the data point at $p=265\,{\rm
MeV}$ from Ref.~\cite{vpiPWA} differs from the fit in Ref.~\cite{NijPW} by more than 
eight standard deviations.

$\epsilon_1$ has a series expansion in $p^2$:
\begin{eqnarray}
  \epsilon_1 &=&  g_1\, p^2 + g_2\, p^4 + g_3\, p^6 + \ldots \,. 
\end{eqnarray}
Fitting this polynomial to the solid line in Fig.~\ref{ep1b} for $7\,{\rm MeV}< p < 
50\,{\rm MeV}$ and weighting low momenta more heavily than high momenta gives 
the values in the first column in Table~\ref{tble_gs}.  To estimate the uncertainty in
the extraction of the $g_i$ we varied the range of momentum and weighting used in 
the fit. The value of $g_1$ is quite stable, while $g_2$ and $g_3$ varied 
by 10\% and 50\% respectively. The effective field theory predictions for the 
coefficients $g_i$ are:
\begin{eqnarray}
  g_1 &=& \frac{2\sqrt{2}}{m_\pi \Lambda_{NN} } \bigg( \frac{8}{15}-
     \frac{\gamma}{m_\pi} \bigg) - \frac{\sqrt{2}}{\Lambda_{NN}^2 } \bigg( 
     \frac{601}{600} -\frac{8}{5}\ln 2 -\frac{5\,\gamma}{m_\pi}+
     \frac{2\,\gamma^2}{m_\pi^2} \bigg) - \frac{\zeta_6}{m_\pi^2} + \frac{8\pi\, \sqrt{2}\, 
      \zeta_2}{M\,\Lambda_{NN}} \,, \nn \\
  g_2 &=& \frac{4\sqrt{2}}{m_\pi^3 \Lambda_{NN} } \bigg( \frac{-32}{35}+
      \frac{5\,\gamma}{3\, m_\pi} \bigg)+ \frac{4\sqrt{2}}{m_\pi^2 \Lambda_{NN}^2} 
     \bigg( \frac{391}{315} -\frac{4}{5}\ln 2 -\frac{589\,\gamma}{120\,m_\pi}+
     \frac{8\,\gamma^2}{3\,m_\pi^2} \bigg) \nn\\*  && 
     +\frac{2\sqrt{2}}{m_\pi^2 \Lambda_{NN}}\, \frac{4\pi}{M} \Big(\zeta_1-
      \frac{10}{3}\zeta_2 \Big)\,, \nn \\
  g_3 &=& \frac{16\sqrt{2}}{m_\pi^5\Lambda_{NN} } \bigg( \frac{16}{21}-
     \frac{7\,\gamma}{5\, m_\pi} \bigg) - \frac{16\sqrt{2}}{m_\pi^4 \Lambda_{NN}^2} 
     \bigg( \frac{241}{200} -\frac{3}{5}\ln 2 -\frac{252409\,\gamma}{50400\,m_\pi}+
     \frac{46\,\gamma^2}{15\,m_\pi^2} \bigg) \nn\\*  && 
     - \frac{8\sqrt{2}}{m_\pi^4 \Lambda_{NN}}\, \frac{4\pi}{M} \Big(\frac{5}{6}\zeta_1-
      \frac{14}{5}\zeta_2 \Big)\,.
\end{eqnarray}
In each $g_i$ the first term is from the order $Q^0$ diagrams in Fig.~\ref{diag_lo},
while the remaining terms are from the order $Q$ diagrams in Fig.~\ref{diag_Q}. 
Using the values in Eq.~(\ref{nums}) gives the predictions in Table.~\ref{tble_gs}.  At
order $Q^0$ the effective field theory is off by a factor of $2$.  The order $Q$
corrections make the predictions closer to the fit values; the error is $\sim
25\%$ for $g_1$ and $g_2$, while $g_3$ is consistent within error.  The effective field
theory is converging onto the experimental $g_i$, but the errors are somewhat larger
than anticipated by the power counting. The convergence for terms in the expansion
of $\epsilon_1$ is faster than the convergence in the $^1S_0$ channel.

{\tighten \begin{table}[!t]
\begin{center} \begin{tabular}{cccccc} 
 & &  Fit to Nijmegen $\epsilon_1$ &  ${\cal O}(Q^0)$ & ${\cal O}(Q)$ & \\ \hline 
& $g_1\, (\rm fm^2)$ & $\: 0.30\pm 0.01$ & $0.55$ & $0.22$ \\
& $g_2\, (\rm fm^4)$ & $-2.0 \pm 0.2$ & $-4.1$ & $-1.5$ \\
& $g_3\, (\rm fm^6)$& $\: 8.7 \pm 4.3$ & $28$ & $9.5$ 
\end{tabular} \end{center} 
{\tighten \caption{Predictions for the coefficients in a momentum expansion of 
$\epsilon_1$ at LO and NLO in the effective field theory. } \label{tble_gs} } 
\end{table} } 

To summarize, we have computed the order $Q^2$ correction to the mixing
parameter ${\epsilon}_1$. The effective theory converges onto the observed
${\epsilon}_1$, and errors are comparable to uncertainties in alternative approaches
where the pion is treated nonperturbatively for $p \sim m_\pi$. When performing low
energy momentum expansions, it is important to use a parameterization of the S
matrix in which the mixing angle has a convergent expansion for $p < m_\pi/2$. The
effective field theory predictions for the coefficients of this expansion converge
towards values extracted from a fit to low energy data. In the future, it will be
interesting to see if including the $\Delta$ or going to one higher order in the $Q$
expansion will provide better agreement for $\epsilon_1$ at $p > m_\pi$.

S.F. was supported in part by NSERC and wishes to thank the Caltech theory group
for their hospitality.  T.M and I.W.S. were supported in part by the Department of
Energy under grant number DE-FG03-92-ER 40701.

Note Added in Proof: While this paper was being reviewed, the authors completed a NNLO
calculation of the phase shifts in the $^1S_0$, $^3S_1$, and $^3D_1$ channels. 
Predictions for the other P and D wave phase shifts were also examined. We find that
in some of these channels the KSW expansion exhibits large corrections at NNLO which
suggest a breakdown of the perturbative treatment of pions.  A detailed discussion can
be found in the preprint \cite{fms2}.

\appendix
\section{T\lc{races in $n$} D\lc{imensions}}

In the standard implementation of dimensional regularization in relativistic
theories, the spin traces are performed in $d$ dimensions \cite{Collins}.  For
non-relativistic nucleon-nucleon scattering the spin traces are often done in
$3$ dimensions, after which the remaining scalar integrals are evaluated in $d=n+1$
dimensions.  This is in agreement with performing a partial wave expansion of the
matrix elements using Clebsh-Gordan coefficients; a procedure specific to $n=3$. 
This approach provides well-defined results for S-wave transitions.  However,
when higher partial waves are considered it becomes necessary to perform the spin
traces in $n$ dimensions.  To see why consider Fig.~\ref{diag_Q}\,e), and replace
the bubble sum by a single $C_0$ for simplicity. The numerator of this graph is
proportional to
\begin{eqnarray}  \label{num}
  \Big( \frac{p^i p^{i'}}{p^2} -\frac{\delta^{ii'}}{n}\Big) {\rm Tr}[ \sigma^i \sigma^j 
    \sigma^m\sigma^{i'} \sigma^{m'} \sigma^{j'} ] \, q^m\, q^{m'}\, k^j\,k^{j'}   \,,
\end{eqnarray}
where $k$ and $q$ are the two loop momenta which run through the pion lines.  First
consider setting $\delta^{ii'}/n=\delta^{ii'}/3$ in Eq.~(\ref{num}) and performing
the trace in $4$ dimensions. At very low momentum, the result can be expanded in
$p/m_\pi$.  When this is done, the amplitude from this graph is proportional to a
constant for low $p$.  However, for a $^3S_1$ to $^3D_1$ transition the amplitude
should be proportional to $p^2$ at low momentum.  The constant indicates that
projection onto $^3S_1-{}^3D_1$ was unsuccessful.  If we keep the $\delta^{ii'}/n$
in Eq.~(\ref{num}), and perform the trace in $3$ dimensions then the amplitude is
still proportional to a constant for low momentum.  However, if the trace in
Eq.~(\ref{num}) is done in $n$ dimensions then the amplitude is proportional to
$p^2$ as it should be.  In the $^3S_1-{}^3D_1$ calculation the two terms in round
brackets in Eq.~(\ref{num}) have $m_\pi^2/\epsilon$ divergences.  These divergences
cancel in the difference no matter how the expression is evaluated, because there is
no operator in this partial wave to absorb an $m_\pi^2/\epsilon$ divergence. 
However, the finite $m_\pi^2$ contributions only cancel when spin traces and
projection operators are evaluated in $n$ dimensions.  Therefore, in this paper all
spin traces will be performed in $n$ dimensions.  

In Ref.~\cite{msw,triton} it was pointed out that the nucleon contact interactions
with no derivatives are invariant under Wigner's SU(4) spin-isospin symmetry for
$a^{(^1S_0)},a^{(^3S_1)} \to \infty$.  If spin traces are performed in $n$
dimensions then it is necessary to treat the isospin traces on the same footing,
otherwise Wigner symmetry will be broken by the regulator.  For this reason,
isospin traces will also be done in $n$ dimensions. For example, if the order
$Q_r^3$ radiation pion calculation in Ref.~\cite{ms2} is performed with spin traces
in $n$ dimensions, but isospin traces in $3$ dimensions then the result is not
proportional to $1/a(^1S_0)-1/a(^3S_1)$.  However, in Ref.~\cite{msw} it was shown
that Wigner symmetry implies that the order $Q_r^3$ graphs should be proportional
to $1/a(^1S_0)-1/a(^3S_1)$.  If all spin and isospin traces are performed in $n$
dimensions then the value of individual order $Q_r^3$ graphs changes, but the sum
gives the same result as in Ref.~\cite{ms2}.  

If the partial wave projection operators are chosen to have the normalization given
in Eq.~(\ref{proj}) then doing the traces in $n$ dimensions does not change any
calculations in the theory without pions.  For S-wave transitions in the theory
with pions this convention amounts to a change of renormalization scheme, since the
difference in evaluating a graph is an overall multiplicative factor of the form $1
+ O(\epsilon)$.  In PDS, subleading terms in the beta functions for coefficients of
four nucleon operators are affected.  When spin and isospin traces are done in $n$
dimensions the NLO $^3S_1$ (or $^1S_0$) amplitude is 
\begin{eqnarray}  \label{Samp}
{ {\cal A}^{(0)} \over [{\cal A}^{(-1)}]\,^2 } &=& -m_\pi^2( \zeta_1 \, \alpha^2 + 
   \zeta_2 ) + {m_\pi^2 g_A^2 \over 2 f^2} \bigg( \frac{M}{4\pi} \bigg)^2
   \Bigg[ {(\hat \gamma^2-\alpha^2) \over 4 \alpha^2} \ln(1+4\alpha^2) -
   {\hat \gamma \over \alpha} \tan^{-1}(2\alpha) \Bigg] \,,
\end{eqnarray}
where $\hat \gamma = \gamma/m_\pi$ and $\zeta_1$ and $\zeta_2$ are given in
Eq.~(\ref{zetas}). Different schemes will give different expressions for
$\zeta_{1,2}$, but the amplitude in Eq.~(\ref{Samp}) will remain the same.  

\section{E\lc{xpressions for} ${\cal Y}$ \lc{and} ${\cal Z}$}

In this appendix we give expressions for ${\cal Y}$ and ${\cal Z}$ which appear in 
Eqs.~(\ref{AmpQ}) and (\ref{Asum}):
\begin{eqnarray}
{\cal Y}(\alpha) &=& -\frac25 + \frac3{10\alpha^2} + \bigg(\frac{3}{8\alpha^5} + 
      \frac{5}{4\alpha^3} - \frac{2\alpha}{5} \bigg) \tan^{-1}(\alpha)  
      -\bigg(\frac{3}{8\alpha^5} +\frac{5}{4\alpha^3} \bigg) \tan^{-1}(2\alpha) \\*
   && + \frac{(15-4\alpha^2)}{80\alpha^6} \ln(1+\alpha^2)  
     - \frac{(3+16\alpha^2+16\alpha^4)}{32\alpha^7} {\rm Im}\,\bigg[  \ply\Big( 
      \frac{2\alpha^2+i\alpha}{1+4\alpha^2} \Big) +  \ply( -2\alpha^2-i\alpha )
      \bigg]  \nn\\*
   && + i \bigg[ \frac{3}{8\alpha^3} +\frac{1}{2\alpha} -\frac{\alpha}{2} - 
      \frac{(3+10\alpha^2)}{16\alpha^5}  \ln(1+4\alpha^2)
      +\frac{(3+16\alpha^2+16\alpha^4)}{128\alpha^7} \ln^2(1+4\alpha^2) \bigg]\,, \nn
   \nn\\ 
{\cal Z}(\alpha) &=& -\frac{7}{40} +\frac{9i}{16\alpha^3}+\frac{21}{40\alpha^2} +
  \frac{3i}{40\alpha}-\frac{3i\alpha}{5} + \frac{29\alpha^2}{200} + 
  \Big(\frac{3\alpha^2}{5} -\frac{9}{16\alpha^4}-\frac{15}{8\alpha^2} \Big) \ln{2} \\*
 && + \frac{3\,(16\alpha^7-50\alpha^3-4i\alpha^2-15\alpha+15i)}{80\alpha^5} 
    \ln(1-i\alpha) \nn\\*
 && + \frac{(-9\,i + 27\,\alpha  - 24\,i\,{{\alpha }^2} + 
    78\,{{\alpha }^3} - 16\,{{\alpha }^5})}{32\alpha^5} \ln(1-2i\alpha) \nn \\*
 && - \frac{(9+48\alpha^2+48\alpha^4)}{64\alpha^6} \bigg[ \frac32 \ln^2(1-2i\alpha)
    + 2 \ply(-1+2i\alpha) + \ply \Big(\frac{1+2i\alpha}{-1+2i\alpha}\Big) +\frac{\pi^2}4
    \bigg]   \nn \,.
\end{eqnarray}
In deriving the formula for ${\cal Z}(\alpha)$ we found it useful to use reduction
formulae due to Tarasov\cite{Tarasov} implemented with the program from
Ref.~\cite{mertig}.

{\tighten

} %end tighten (references & figure captions)

\end{document}